\documentstyle[preprint,aps,epsfig]{revtex}
\preprint{USM-TH-101}
\begin{document}
\title{Remarks on Screening in a Gauge-Invariant Formalism}
\author{ Patricio Gaete \thanks{E-mail: pgaete@fis.utfsm.cl},
and Iv\'an Schmidt \thanks{E-mail: ischmidt@fis.utfsm.cl}}
\address{Departamento de F\'{\i}sica, Universidad T\'ecnica F.
Santa Maria, Valpara\'{\i}so, Chile} \maketitle

\begin{abstract}
In this paper we display a direct and physically attractive
derivation of the screening contribution to the interaction
potential in the Chiral Schwinger model and generalized
Maxwell-Chern-Simons gauge theory. It is shown that these results
emerge naturally when a correct separation between gauge-invariant
and gauge degrees of freedom is made. Explicit expressions for
gauge-invariant fields are found.
\end{abstract}
\smallskip

PACS number(s): 12.20.Ds, 11.15.Tk

\section{INTRODUCTION}

The binding energy of an infinitely heavy quark-antiquark pair
represents a fundamental concept which is expected to play an
important role in the understanding of non-Abelian theories and
especially of quark confinement. In fact, a linearly increasing
quark-antiquark pair static potential provides the simplest
criterion for confinement, although unfortunately there is up to
now no known way to analytically derive the confining potential
from first principles. However, as it is well known, in certain
theories the rising potential can be screened at large distances
by dynamical charges. An illustrative example arises when one
considers two dimensions of spacetime, where it was shown in Ref.
\cite{Gross,Elcio} that matter fields can screen any Abelian
fractional charge. In three dimensions massive fermions screen any
charge \cite{Elcio2}. In this context it may be recalled the
confining and screening nature of the potential in QED3
\cite{Banerjee,Gaete1}. As the three dimensional fermions induce a
topological mass for the photon, the logarithmically rising
potential is transformed into an exponentially decreasing one. We
further note that recently the stability of the above potential in
the presence of a self-interaction among fermions has been studied
\cite{Ghosh}. In particular it was considered the Thirring
interaction. As a result it was argued that when the Thirring
coupling $g$ is positive, the new model displays a marked
departure of a qualitative nature from the results of  Ref.
\cite{Banerjee} at short distances. More precisely, the resulting
potential obtained has the form of a well, which may be contrasted
with the result of \cite{Banerjee,Gaete1} where for small
separation the potential tends to a logarithmic Coulomb potential.
In this way, the author of Ref. \cite{Ghosh} is lead to the
conclusion that the effect of adding the Thirring term is to
stabilize the charge system. Despite their relevance, this study
was carried out in a gauge fixed scheme, and we think that their
results should be confirmed by a gauge independent analysis.

Meanwhile, in a previous paper \cite{Gaete1} we have proposed a
general framework for studying the confining and screening nature
of the potential in QED3 in terms of gauge-invariant but
path-dependent field variables. According to this formalism, the
interaction potential between two static charges is obtained once
a judicious identification of the physical degrees of freedom is
made. This procedure leads to the physical phenomena of
electrostatic and dressed electrons, where we refer to the cloud
made out of the vector potentials around the fermions as dressing.
In this sense, our formalism has provided a method for the
determination of the potential between charges which, in our view,
is of interest both for its simplicity and physical content. We
also point out that a similar analysis has been developed for the
Schwinger model \cite{Gaete2}.

In this Brief Report we will continue our program
\cite{Gaete1,Gaete2} to study the structure of the interaction
energy between charged fermions. In the next section we develop
further the discussion that was begun in \cite{Gaete2} by
considering how the charged fermions behave in the Chiral
Schwinger model \cite{Schmidt}. Section III is concerned with the
calculation of the interaction energy in QED3 with a Thirring
interaction term among fermions. Particular care is paid to
reexamine the consequences of including this term in the confining
and screening nature of the potential.

\section{PRELIMINARY: GAUGE-INVARIANT VARIABLES FORMALISM}

First, let us briefly review the framework of the gauge-invariant
but path-dependent field variables formalism as described in Ref.
\cite{Gaete0,Gaete1,Gaete2}. Accordingly, we consider the
gauge-invariant field
\begin{equation}
{\cal A}_{\mu }(y)=A_{\mu }(y)+\partial _{\mu }\Lambda (y) .
\label{bia}
\end{equation}
The function $ \Lambda (y) $ is defined by
\begin{equation}
\Lambda (y)=-\int_{C_{\xi y}}dz^{\nu }A_{\nu }(z), \label{fer}
\end{equation}
where the path integral is to be evaluated along some contour
$C_{\xi y\text{ }}$ connecting $\xi $ and $y$. Here $A_{\mu }$ is
the usual electromagnetic potential and, in principle, it is taken
in an arbitrary gauge. It can easily be verified that ${\cal
A}_{\mu }(y)$ is invariant with respect to gauge transformations $
 A_\mu  \left( y \right) \mapsto A_\mu  \left( y \right) +
\partial _\mu  \Theta \left( y \right)$. It is now important to
notice that the gauge invariant field (\ref{bia}) depends not only
on the points  $\xi $ and $y$ but also on the path. Furthermore,
by choosing a spacelike path from the point $\xi^k $ to $y^k$, on
a fixed time slice, it is possible to express the gauge-invariant
field in terms of the magnetic (${\bf B}$) and electric (${\bf
E}$) fields as:
\begin{equation}
{\cal A}_0 \left( {t,{\bf y}} \right) =  - {\bf y} \cdot
\int\limits_0^1 {d\alpha } {\bf E}\left( {t,\alpha {\bf y}}
\right), \label{lag3}
\end{equation}
\begin{equation}
{\bf {\cal A}}\left( {t,{\bf y}} \right) =  - {\bf y} \wedge
\int\limits_0^1 {d\alpha } \alpha {\bf B}\left( {t,\alpha {\bf y}}
\right) , \label{lag4}
\end{equation}
where  $\alpha$  $\left( {0 \le \alpha  \le 1} \right)$ is the
parameter describing the contour $ y^k  = \xi ^k  + \alpha \left(
{y - \xi } \right)^k$ with $k=1,2,3$. For simplicity we have
assumed the reference point $\xi$ at the origin. The above
expressions coincide with the Poincar\'{e} gauge conditions
\cite{Gaete0} for the path-dependent fields ${\cal A}_{\mu}$,
while other contour choices coincide with other gauge conditions
for these fields. For reasons that will become evident later, we
now focus our attention on the fermion field. In the context of
our formalism, the charged matter field together with the
electromagnetic cloud (dressing) which surrounds it, is given by
\cite{Gaete0,Dirac},
\begin{equation}
\Psi (y) = \exp \left( { - ie\int_{C_{\xi y} } {dz^\mu  A_\mu (z)}
} \right)\psi (y) . \label{expo1}
\end{equation}
Thanks to our path choice, the physical fermion (\ref{expo1}) then
becomes
\begin{equation}
\Psi (y) = \exp \left( { - ie\int_0 ^y {dz^k A_k (z)} }
\right)\psi (y) . \label{expo2}
\end{equation}
The expressions (\ref{lag3}), (\ref{lag4}) and (\ref{expo2}) will
form the basis of our subsequent considerations, where the gauge
invariance of the formalism guarantees that the relevant physical
information will be preserved. We conclude this brief introduction
to gauge invariant variables by pointing out that the breaking of
the gauge invariance of the fields in the standard formalism is
transformed into breaking of the translational invariance in the
path-dependent formalism. This drawback is avoided by letting the
reference point $ \xi^k $ go to infinity.

As already mentioned, we now want to study the interaction energy
between external probe sources in the Chiral Schwinger model (CSM)
\cite{Schmidt}, which consists of a U(1) gauge field coupled to
chiral fermions in two-dimensional spacetime. As was shown by
Jackiw-Rajaraman \cite{JRaj} this theory can be consistently
quantized and the quantum theory is unitary, in spite of its gauge
anomaly. Faddeev and Shatashvili \cite{FShat} have suggested a
modification of the canonical quantization by addition new degrees
of freedom through a Wess-Zumino action. At present two formalisms
of the Chiral Schwinger model are available: the gauge
noninvariant one \cite{JRaj} and the gauge invariant one
\cite{FShat,FalK,Girotti}.

We begin by recalling the bosonized form of the gauge-invariant
version under consideration \cite{Schmidt}:
\begin{equation}
\begin{array}{r}
{\cal L} =  - \frac{1}{4}F_{\mu \nu }^2  + \frac{1}{2}\left(
{\partial _\mu  \varphi } \right)^2  - \frac{g}{{\sqrt \pi
}}\left( {g^{\mu \nu }  - \varepsilon ^{\mu \nu } }
\right)\partial _\mu  A_\nu  \varphi  + \frac{{ag^2 }}{{2\pi
}}A_\mu  A^\mu    \\ + \frac{1}{2}\left( {a - 1} \right)\left(
{\partial _\mu  \theta } \right)^2  - \frac{g}{{\sqrt \pi }}\left[
{\left( {a - 1} \right)g^{\mu \nu }  + \varepsilon ^{\mu \nu } }
\right]  \partial _\mu  A_\nu  \theta  - A_0 J^0,  \\
\end{array} \label{bosoni}
\end{equation}
where $J^0$ is the external current. The parameter $a$ reflects
the ambiguity of fermionic radiative corrections or may be
considered as a bosonization ambiguity parameter. Here we focus
our attention to the $a>1$ case. As we have already indicated in
\cite{Gaete2}, to compute the interaction energy we need to carry
out the integration over $\varphi$ and $\theta$ in
Eq.(\ref{bosoni}). Once this is done, we arrive at the following
effective Lagrangian for the gauge fields
\begin{equation}
{\cal L} =  - \frac{1}{4}F_{\mu \nu } \left( {1 + \frac{{g^2
}}{\pi }\frac{a}{{a - 1}}\frac{1}{{\partial ^2 }}} \right)F^{\mu
\nu }  - A_0 J^0 .  \label{bosoni2}
\end{equation}
One immediately sees that this expression is very similar to the
effective Lagrangian for the Schwinger model. Notwithstanding, in
order to put our discussion into context it is useful to summarize
the relevant aspects of the analysis described previously
\cite{Gaete2}. Thus, our first undertaking is to calculate the
expectation value of the Hamiltonian in the physical state $
|\Omega\rangle $, which we will denote by $\langle
H\rangle_{\Omega}$.

We now proceed to obtain the Hamiltonian. For this we restrict our
attention to the Hamiltonian framework of this theory. The
canonical momenta read $\Pi ^\mu  = \left( {1 + \frac{{g^2 }}{\pi
}\frac{a}{{a - 1}}\frac{1}{{\partial ^2 }}} \right)F^{\mu 0}$, and
one immediately identifies the sole primary constraint $\Pi^0=0$.
The canonical Hamiltonian is given by
\begin{equation}
H_C  = \int {dx\left( { - \frac{1}{2}\Pi _1 \left( {1 + \frac{{g^2
}}{\pi }\frac{a}{{a - 1}}\frac{1}{{\partial ^2 }}} \right)^{ - 1}
\Pi ^1  + \Pi ^1 \partial _1 A_0  + A_0 J^0 } \right)}.
\label{hamil}
\end{equation}
With this at hand, the consistency condition $ {\dot \Pi}_0 = 0$
lead to the secondary constraint
\begin{equation}
\Omega _1 \left( x \right) = \partial _1 \Pi ^1  - J^0 ,
\label{const}
\end{equation}
and the time stability of the secondary constraint does not induce
further constraints. Therefore, in this case there are two
constraints, which are first class. The extended Hamiltonian that
generates translations in time then reads
\begin{equation}
H = H_C  + \int {dx\left( {c_0 (x)\Pi _0 (x) + c_1 (x)\Omega _1
(x)} \right)} , \label{canon}
\end{equation}
where $ c_0(x) $ and $ c_1(x) $ are the Lagrange multiplier
fields. Moreover, it follows from this Hamiltonian that $ \mathop
{A_0 }\limits^ \cdot  \left( x \right) = \left[ {A_0 \left( x
\right),H} \right] = c_0 \left( x \right) $ , which is an
arbitrary function. Since $ \Pi^0 = 0$ always, neither $ A^0 $ nor
$ \Pi^0 $ are of interest in describing the system and may be
discarded from the theory. Thus the Hamiltonian takes the form
\begin{equation}
H = \int {dx\left\{ { - \frac{1}{2}\Pi _1 \left( {1 + \frac{{g^2
}}{\pi }\frac{a}{{a - 1}}\frac{1}{{\partial ^2 }}} \right)^{ - 1}
\Pi ^1  + c^{\prime} (x)\left( {\partial _1 \Pi ^1  - J^0 }
\right)} \right\}} , \label{canon1}
\end{equation}
where $ c^{\prime }(x)=c_{1}(x)-A_{0}(x)$.

Now the presence of the arbitrary quantity $ c^{\prime }(x)$ is
undesirable since we have no way of giving it a meaning in a
quantum theory. To avoid this trouble we introduce a supplementary
condition on the vector potential such that the full set of
constraints becomes second class. A particularly convenient
condition is
\begin{equation}
\Omega _2 (x) = \int_0^1 {d\alpha x^1 } A_1 (\alpha x) = 0 ,
\label{poin}
\end{equation}
where, as before, $\alpha$ is the parameter describing a spacelike
straight line of integration. In this case, the basic Dirac
brackets between the canonical variables have the following form:
\begin{equation}
\left\{ {A_1 (x),A^1 (y)} \right\}^ *   = 0 = \left\{ {\Pi _1
(x),\Pi ^1 (y)} \right\}^ * , \label{paren}
\end{equation}
\begin{equation}
\left\{ {A_1 (x),\Pi ^1 (y)} \right\}^ *   = \delta ^{(1)} (x - y)
- \partial _1^x \int_0^1 {d\alpha x^1 \delta ^{(1)} } (\alpha x -
y) . \label{paren2}
\end{equation}
Next, as remarked by Dirac \cite{Dirac}, the physical states $
|\Omega\rangle $ correspond to the gauge invariant ones. In this
way, we consider the stringy gauge-invariant $ \mid
\overline{\Psi} (y)\Psi (y ^{\prime})\rangle $ state,
\begin{equation}
\mid \Omega  \rangle  \equiv \mid \overline{\Psi} (y)\Psi (y
^{\prime})\rangle  =\overline{\psi }(y)\exp \left( { -
iq\int_y^{y^{\prime}} {dz^1 A_1 (z)} } \right)\psi (y)\left| 0
\right\rangle , \label{ener3}
\end{equation}
where $ | 0\rangle $ is the physical vacuum state.

We have finally assembled the tools to determine the interaction
energy. Recalling again that the fermions are taken to be static,
we can therefore substitute $ \partial^2$ by -$\partial^2_1 $ in
the Hamiltonian. As a consequence, the expectation value $\langle
H\rangle_{\Omega}$ simplifies to
\begin{equation}
\left\langle H \right\rangle _\Omega   = \left\langle \Omega
\right|\int {dx_1 } \left( { - \frac{1}{2}\Pi _1 \left( {1 -
\frac{{g^2 }}{\pi }\frac{a}{{a - 1}}\frac{1}{{\partial _1^2 }}}
\right)^{ - 1} \Pi ^1 } \right)\left| \Omega  \right\rangle .
\label{scree1}
\end{equation}
Taking into account the above Hamiltonian structure, the resulting
interaction energy of the dressed fermion-antifermion system takes
the form
\begin{equation}
\left\langle H \right\rangle _\Omega   = \left\langle H
\right\rangle _0  + \frac{{q^2 \sqrt \pi  }}{{2g}}\sqrt {\frac{a -
1}{{a}}} \left( {1 - e^{ - \frac{g}{{\sqrt \pi }}\sqrt
{\frac{a}{{a - 1}}} |y - y^\prime  |} } \right), \label{scree2}
\end{equation}
where $ \left\langle H \right\rangle _o  = \left\langle 0
\right|H\left| 0 \right\rangle $ . The second term on the
right-hand side of Eq.(\ref{scree2}) is clearly dependent on the
distance between the external static fields. Therefore the
potential for two opposite charges located at $y$ and $y^{\prime}$
is given by
\begin{equation}
V = \frac{{q^2 \sqrt \pi  }}{{2g}}\sqrt {\frac{a - 1}{{a}}} \left(
{1 - e^{ - \frac{g}{{\sqrt \pi  }}\sqrt {\frac{a}{{a - 1}}} |y -
y^\prime  |} } \right). \label{scree3}
\end{equation}
This result agrees with that of Ref.\cite{Berger}, and finds here
an independent derivation. An immediate consequence of the
Eq.(\ref{scree3}) is that it saturates at $ \frac{{q^2 \sqrt \pi
}}{{2g}}\sqrt {\frac{a - 1}{{a}}} $ for large $ |y - y^\prime| $.
In other terms, the charged fermions behave in the same way as in
the Schwinger model, that is, the probe charges are screened. The
point we wish to emphasize, however, is that within this framework
one circumvents completely the need to introduce the Wilson loop,
where subtleties related to the correct calculation must be
considered \cite{Hag,Fal}. Thus, in the context of our present
formalism, the potential energy between fermions can be directly
obtained once the structure of the electromagnetic cloud around
static fermions is known.

We also observe at this stage that the gauge-invariant variables
${\cal A}_{\mu }$ commute with the sole first class constraint
(Gauss law), corroborating the fact that these fields are physical
variables \cite{Dirac}. This last point enables us to arrive at
the result (\ref{scree3}) by an alternative but equivalent way. To
this end we first note that the physical electron ( i.e. an
electron together with the electric field surrounding it ), Eq.
(\ref{expo1}), may be rewritten as
\begin{equation}
\Psi \left( y \right) = \exp \left( { - ie\int\limits_0^1 {dz^1
A_1^L \left( z \right)} } \right)\psi \left( z \right) ,
\end{equation}  \label{modif}
where $ A_1^L $ refers to the longitudinal part of $ A_1 $. It is
worth noting here that the above expression uses a modified form
for the electromagnetic cloud in the Poincar\'e gauge Eq.
(\ref{expo2}), which is equivalent to the Coulomb gauge
\cite{GaeteG}. Having made this observation and from the previous
Hamiltonian analysis, we can write immediately the following
physical scalar potential
\begin{equation}
{\cal A}_0 \left( {t,x} \right) = \int\limits_0^1 {d\alpha x^1
E_1^L \left( {t,\alpha x} \right)} = \int\limits_0^1 {d\alpha x^1
\left( {1 - \frac{{g^2 }}{\pi }\frac{a}{{a - 1}}\frac{1}{{\partial
_1^2 }}} \right)} _{\alpha x}^{ - 1} \frac{{\partial _1^{\alpha x}
\left( { - J^0 \left( {\alpha x} \right)} \right)}}{{\partial
_{\alpha x}^2 }} , \label{modifi}
\end{equation}
where $J^0$ is the external current. The static current describing
two opposites charges $q$ and $-q$ located at $y$ and $y^{\prime}$
is then described by
\begin{equation}
J^0 (t,x)= q \{ \delta(x-y)-\delta(x-y^{\prime}) \}.
\label{fuente}
\end{equation}
Substituting this back into the Eq. (\ref{modifi}), we obtain
\begin{equation}
V = q\left( {{\cal A}_0 \left( y \right) - {\cal A}_0 \left(
{y^\prime } \right)} \right) = \frac{{q^2 \sqrt \pi }}{{2g}}\sqrt
{\frac{a - 1}{{a}}} \left( {1 - e^{ - \frac{g}{{\sqrt \pi }}\sqrt
{\frac{a}{{a - 1}}} |y - y^\prime  |} } \right). \label{pot2}
\end{equation}
It is clear from this discussion that a correct identification of
physical degrees of freedom is a key feature for understanding the
physics hidden in gauge theories. According to this viewpoint,
once that identification is made, the computation of the potential
is achieved by means of Gauss law \cite{Hag}.

\section{GENERALIZED MAXWELL-CHERN-SIMONS GAUGE THEORY}

We now extend the analysis of the previous section to a
(2+1)-dimensional topologically massive gauge theory, which
includes a self-interaction among fermions, the so-called
generalized Maxwell-Chern-Simons gauge theory \cite{Ghosh}. As
mentioned above, in this section we concentrate on the effect of
including the self-interaction term in the confinement and
screening nature of the potential.

Before going to the derivation of the interaction energy, we will
describe very briefly the model under consideration. It is
described by the following Lagrangian:
\begin{equation}
 {\cal L} =  - \frac{{pe^2 }}{4}F_{\mu \nu } F^{\mu \nu }  +
\frac{{qe^2 }}{2}\varepsilon _{\mu \nu \lambda } A^\mu F^{\nu
\lambda }  + \overline \psi  i\gamma ^\mu \left( {\partial _\mu -
ieA_\mu  } \right)\psi  \\ - m\overline \psi  \psi  +
\frac{g}{2}|\overline \psi \gamma ^\mu  \psi |^2  + J_0 A^0  \\ ,
\label{gohsh1}
\end{equation}
where $ F_{\mu\nu} = \partial_{\mu}A_{\nu}-\partial_{\nu}A_{\mu}$,
$J^0$ is an external current and $ p = \frac{1}{e^{2}}$, $ q =
\frac{\chi}{2e^{2}}$.

Next, in order to linearize this theory, we introduce the
auxiliary field $ B_{\mu} $. It follows that the expression
(\ref{gohsh1}) can be rewritten as
\begin{equation}
{\cal L} =  - \frac{{pe^2 }}{4}F_{\mu \nu }^2  + \frac{{qe^2
}}{2}\varepsilon _{\mu \nu \lambda } A^\mu  F^{\nu \lambda }  +
\overline \psi  i\gamma ^\mu  \left( {\partial _\mu   - ieA_\mu -
iB_\mu  } \right)\psi  \\
  - \frac{1}{{2g}}B_\mu ^2  - m\overline \psi  \psi  + J^0 A_0  \\
  . \label{gohsh2}
\end{equation}
By using bosonization methods, it can be shown that in the large
fermion mass limit, the Lagrangian (\ref{gohsh2}) then becomes
${\cal O}(\frac{1}{m})$ \cite{Ghosh}:
\begin{equation}
 {\cal L} =  - \frac{a}{4}W_{\mu \nu } W^{\mu \nu }  + \frac{\alpha }
{2}\varepsilon _{\mu \nu \lambda } W^\mu  W^{\nu \lambda }  -
\frac{1}{{2g}}B_\mu ^2  - \frac{{pe^2 }}{4}F_{\mu \nu } F^{\mu \nu
}  \\ + \frac{{qe^2 }}{4}\varepsilon _{\mu \nu \lambda } A^\mu
F^{\nu \lambda }  \\ ,\label{gohsh3}
\end{equation}
where $W_{\mu} = B_{\mu} + eA_{\mu}$, $W_{\mu\nu} =
\partial_{\mu}W_{\nu} - \partial_{\nu}W_{\mu}$, $\alpha = -\frac{1}{8\pi}$
and $a = -\frac{1}{8{\pi}m}$.

Following our earlier discussion, this expression allows us to
derive an effective Lagrangian. Thus, after the $B_{\mu}$ fields
are integrated away, one gets up to ${\cal O}(\frac{1}{m})$ a
generalized Maxwell-Chern-Simons gauge theory:
\begin{equation}
{\cal L} =  - \frac{1}{4}F_{\mu \nu } \left( {1 + e^2 a\left[ {1 -
12\alpha ^2 g^2 \partial ^2 } \right]} \right)F^{\mu \nu }  \\ +
\frac{\chi }{2}\varepsilon _{\mu \nu \lambda } A^\mu
\partial ^\nu A^\lambda
\\ .\label{electri}
\end{equation}

It is now once again straightforward to apply the gauge-invariant
formalism discussed in the preceding section. For this purpose, we
start by observing that the canonical momenta read $\Pi^{\mu} = -
\left( {1 + e^2 a\left[ {1 - 12\alpha ^2 g^2
\partial ^2 } \right]} \right) F^{0\mu }  + \frac{\chi }{2}
\varepsilon ^{0 \mu \nu } A_\nu$. As we can see there is one
primary constraint $\Pi^{0} = 0$, and  $\Pi^{i} =  \left( {1 + e^2
a\left[ {1 - 12\alpha ^2 g^2 \partial ^2 } \right]} \right) E^{i}
+ \frac{\chi }{2} \varepsilon ^{i j } A_j$  $(i,j=1,2)$. The
canonical Hamiltonian for this system is in this case
\begin{equation}
H_C  = \int {d^2 } x\left\{ \begin{array}{r}
  - \frac{1}{2}F_{i0} \left( {1 + e^2 a\left[ {1 - 12\alpha ^2 g^2
\partial ^2 } \right]} \right) F^{i0}  + \frac{1}{4}F_{ij}
\left( {1 + e^2 a\left[ {1 - 12\alpha ^2 g^2
\partial ^2 } \right]} \right)  F^{ij} \\ +
\Pi ^i \partial _i A_0  - \frac{\chi}{2} \varepsilon ^{ij} A_0
\partial _i A_j + A_0 J^0  \\\end{array} \right\} .
\label{electri2}
\end{equation}
The conservation in time of the constraint $\Pi^0$ leads to the
secondary constraint (Gauss law)
\begin{equation}
\Omega _1 \left( x \right) = \partial _i \Pi ^i  + \frac{\chi}{2}
\varepsilon _{ij} \partial ^i A^j  - J^0  = 0 . \label{electri3}
\end{equation}
There are no more constraints in the theory and the two we have
found are first class. It follows, therefore, that the total
Hamiltonian (first class) is given by
\begin{equation}
H = H_C  + \int {d^2 } x\left( {c_0 \left( x \right)\Pi _0 \left(
x \right) + c_1 \left( x \right)\Pi _1 \left( x \right)} \right) ,
\label{electri4}
\end{equation}
where we recall from Eq.  (\ref{canon}) that $ c_0(x) $ and $
c_1(x) $ are arbitrary functions. We also recall that $A_0$ is not
a dynamical variable. Thus the total Hamiltonian $H$ is given as
\begin{equation}
H = \int {d^2 x\left\{ { - \frac{1}{2}F_{i0} F^{i0}  +
\frac{1}{4}F_{ij} F^{ij}  + c^\prime  \left( x \right)\left[
{\partial _i \Pi ^i  + \frac{\chi}{2}\varepsilon
^{ij}\partial_{i}A_j - J^0 } \right]} \right\}} , \label{ham}
\end{equation}
where $ c^{\prime }(x)=c_{1}(x)-A_{0}(x)$.

As before, we now proceed to impose a supplementary condition on
the vector potential such that the full set of constraints becomes
second class. Therefore, we once again write the supplementary
condition as
\begin{equation}
\Omega _2 (x) = \int_0^1 {d\alpha x^i } A_i (\alpha x) = 0 ,
\label{fixing}
\end{equation}
where $\alpha$ is the parameter describing a spacelike straight
line of integration. Correspondingly, the fundamental Dirac
brackets are given by
\begin{equation}
\left\{ {A_i (x),A^j (y)} \right\}^ *   = 0 = \left\{ {\Pi _i
(x),\Pi ^j (y)} \right\}^ * , \label{parent1}
\end{equation}
\begin{equation}
\left\{ {A_i (x),\Pi ^j (y)} \right\}^ *   = \delta_i^j \delta
^{(2)} (x - y) - \partial _i^x \int_0^1 {d\alpha x^j \delta ^{(2)}
} (\alpha x - y) . \label{parent2}
\end{equation}

We now have all the information required to compute the  potential
energy for this theory. As we pointed above, this calculation is
straightforward by using
\begin{equation}
{\cal A}_{0}(t,{\bf x)=}\int_{0}^{1}d\alpha \text{ }%
x^{i}E_{i}^{L}(t,\alpha {\bf x}),  \label{ener}
\end{equation}
which may be rewritten as
\begin{equation}
{\cal A}_0 \left(t,{\bf x} \right) = \int\limits_0^1 {d\alpha
\frac{1}{{1 + e^2 a\left( {1 - 12\alpha ^2 g^2 \partial ^2 }
\right)}}} \frac{{x^i \partial _i^{\alpha x} \left( { - J^0 \left(
{\alpha x} \right)} \right)}}{{\nabla _{\alpha x}^2  - \chi ^2 }}.
\label{ener2}
\end{equation}
Accordingly, for $ J^0(t,{\bf x}) = q\delta^{(2)}({\bf x},{\bf
a})$, expression (\ref{ener2}) takes the form
\begin{equation}
{\cal A}_0 \left(t,{\bf x} \right) = \frac{q}{{2\pi }}\frac{1}{{1
+ e^2 a\left( {1 - 12\alpha ^2 g^2 \partial ^2 } \right)}}\left(
{K_0 \left( {\chi |{\bf x} - {\bf a}|} \right) - K_0 \left( {\chi
|{\bf a}|} \right)} \right). \label{ener3}
\end{equation}
By means of Eq. (\ref{ener3}) we evaluate the potential energy for
a pair of static pointlike opposite charges at ${\bf y}$ and ${\bf
y^{\prime}}$, that is,
\begin{equation}
V = q\left( {{\cal A}_0 \left( {\bf y} \right) - {\cal A}_0 \left(
{{\bf y}^\prime } \right)} \right) =  - \frac{{q^2 }}{\pi
}\frac{1}{{1 + e^2 a\left( {1 - 12\alpha ^2 g^2 \partial ^2 }
\right)}}K_0 \left( {\chi |{\bf y} - {\bf y}^\prime  |} \right)
.\label{ener4}
\end{equation}
Considering again that the fermions are taken to be static, we can
write immediately the potential energy as
\begin{equation}
V =  - Q^2 K_0 \left( {\chi |{\bf y} - {\bf y}^\prime  |} \right)
- \frac{\beta }{\eta }Q^2 \nabla ^2 K_0 \left( {\chi |{\bf y} -
{\bf y}^\prime |} \right) , \label{ener5}
\end{equation}
where $Q\equiv\frac{q^{2}}{\pi\alpha}$,
$\beta=-12e^{2}a\alpha^{2}g^{2}$, $\eta=1+e^{2}a$ and $K_0$ is a
modified Bessel function. This result explicitly shows the effect
of including the self-interaction term, in the form of the second
term on the right-hand side of Eq.(\ref{ener5}). It is
straightforward to see that when $g=0$ the potential (\ref{ener5})
reduces to the one in \cite{Banerjee,Gaete1}, as indeed it should.
Now we recall that the limiting form of $K_0$ for small $\chi r$
($r\equiv|{\bf y} - {\bf y}^\prime|$) is $ K_0 \sim - \ln \left(
{\chi r} \right)$, and hence expression (\ref{ener5}) reduces to
\begin{equation}
V = Q^2 \left( {\ln z - \gamma \frac{1}{{z^2 }}} \right),
\label{ener6}
\end{equation}
where $z\equiv\chi r$  and $\gamma  \equiv \frac{{12e^2 a\alpha ^2
g^2 \chi ^2 }}{{1 + e^2 a}}$, which is a rather small number since
$a\sim\frac{1}{m}$. In this way one obtains that in the short
distances regime the potential has the form of a well (Fig.1), as
was noted in Ref.\cite{Ghosh}. Figure 1 represents the potential
energy Eq.(\ref{ener6}) for the case of $\gamma=0.01$, plotted
with respect to $z$. It is important to notice that the presence
of the second term on the right-hand side of Eq.(\ref{ener6}),
which predominates for small $z$ values, causes $V$ to be
attractive at short distances. One is thus lead to the interesting
conclusion that the effect of adding the self-interaction term is
to generate stable bound states of quark-antiquark pairs at short
distances. However, the central difference between the above
analysis and that of Ref. \cite{Ghosh} lies on the fact that the
potential Eq.(\ref{ener5}) is present at the classical level too.
In this context, the present gauge-invariant investigation
supplements the earlier analysis done in  Ref. \cite{Gaete1}, as
well as it reveals the general applicability of the
gauge-invariant approach.
\begin{figure}
\begin{center}
\epsfig{figure=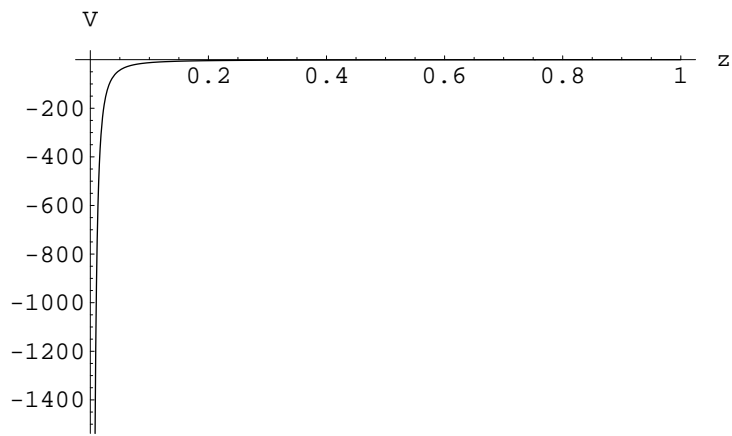 , height=6cm} \caption{Shape of the
potential, Eq.(40), at short distances.}
\end{center}
\end{figure}

\section{ACKNOWLEDGMENTS}

Work supported in part by Fondecyt (Chile) grant 1980149 and grant
1000710, and by a C\'atedra Presidencial (Chile).

\end{document}